Changing the Scene: applying four models of social evolution to the scenescape

Daniel Silver
Thiago Silva
Patrick Adler

**Abstract:** This paper elaborates a multi-model approach to studying how local scenes change. We refer to this as the "4 D's" of scene change: development, differentiation, defense, and diffusion. Each posits somewhat distinct change processes, and has its own tradition of theory and empirical research, which we briefly review. After summarizing some major trends in scenes and amenities in the US context, for each change model, we present some initial findings, discussing data and methods throughout. Our overall goal is to point toward new research arcs on change models of scenes, and to give some clear examples and directions for how to think about and collect data to understand what makes some scenes change, others not, why, and in what directions.

**Scenes research: the current situation**

The scenes perspective on social research grows out of a perception that cities and communities around the globe have been undergoing profound and dramatic changes (D. A. Silver and Clark 2016; Navarro et al. 2012; Wu et al. 2021). While the precise form and pace of change varies, there are nevertheless common motifs. Most centrally, the salience of culture and consumption is on the rise, along with an expanding role for a host of amenities in defining the meaning, opportunities, and value-propositions offered by cities and neighborhoods (Klekotko 2020; Mateos-Mora, Navarro-Yañez, and Rodríguez-García 2021; Daniel Silver 2012).

More of the economy is geared toward providing consumer and leisure services, and aesthetics and design pervade aspects of nearly all products. Various forms of cultural participation, both formal and informal, are rising (da Silva, Clark, and Cabaço 2014). Where to live, and near whom, is increasingly bound up with the amenities available in a place, such as parks, restaurants, entertainment venues, cafes, sports, distinctive shops, and more. The shift for many to working from home only exacerbates this trend; it provides new freedom for individuals and families to make residential decisions based upon the style of life provided by a locale, and to re-imagine their own neighborhoods from a new vantage point that mixes leisure and working hours. Accordingly, political debate and public policy increasingly focus on style of life issues, as targets and focal points for conflict, contestation, identity, and organization (Grodach and Silver 2012; Miller and Silver 2015). Dozens if not hundreds of papers and books by scenes researchers and related work document aspects of these changes in many cities across the globe (Wu 2013, Jeong 2019, Knudsen, Clark and Silver 2016, Navarro and Rodríguez-García, 2014; Zapata et al., 2020).

We developed the theory of scenes from a wide-ranging dialogue with many international collaborators to make the crystallization of these processes evident and legible for social science research. In general terms, the theory holds that more and more cities and communities worldwide are best described not so much in terms of their landscapes, but their scenescapes: the overall style or scene generated by the joint operation of their physical environment and its aesthetics (architecture, greenery, etc); the people who reside, work, and frequent them (including their demographic and socio-economic characteristics); the activities they support (playing saxophone or dancing in a park, drinking coffee in a coffee shop, tending a public garden, etc); and the values and symbolic meanings they evoke (such as personal self-expression, local authenticity, communitarian neighborliness, or glamorous theatricality).

While we and others elaborate the specifics of these ideas in detail elsewhere (Navarro Yáñez 2013; Daniel Silver, Clark, and Navarro Yanez 2010, Klekotko and Navarro 2015), the guiding ambition throughout is integrative. We join traditional social science concepts like tradition, ethnicity, legitimacy, or charisma -- as well as standard local area measures such as ethnic composition, income, age, or voting – with newer or less discussed concepts like glamour, transgression, self-expression, theatricality, and authenticity. The goal is not to replace but to join and combine, precisely because the social world itself changes through accretion of new or small activities and values into old and larger formations.

Many other and past analyses discuss aspects of key scenes concepts, but often in a generic or residual way. For example, concepts like "post-industrial" or "consumer society" point toward "something else" after or in addition to industrialism or producer-oriented social arrangements. But they say little about what this "something else" consists in. The theory of scenes aims to develop concepts, methods, and measurements to fill in this "something else" with substance and content: not consumption in general, but specific consumption practices, which become more refined and specialized as consumption become more central; not a generic idea of work proceeding outside of factories and plants, but specific investigations of how work, residence, and politics are reorganized in the light of expanding salience of design, aesthetics, consumption, amenities, and scenes.

However, while scenes concepts and research emerges from an analytical impulse to respond to major societal changes, very rarely in practice have scenes researchers examined how and why scenes change. Instead, the vast bulk of scenes research has been synchronic. It aims to characterize the types of scenes across many locales at a single time point. While research often connects the type of scene in a place to subsequent changes in local economies, residential patterns, and political orientations, the scene in these studies is presumed to be itself relatively stable (see esp. ch. 7 in D. A. Silver and Clark 2016).

This assumption of stability was always made for analytical convenience, but it arose from understandable reasons. The first concern data available when the theory of scenes was developed. At that time, data on amenities was difficult to come by, and was generally only available at a single time point (or combining years was extremely difficult and labor-intensive). Since data and concepts tend to grow together, with primarily synchronous and geographically comparative data to hand, we tended to develop our ideas with a view toward capturing what scene is present in a given location and rendering many locations comparable to each other. While there is dynamism here, it is about dynamic relations among people, places, activities, values, and subsets of these. But it is not dynamic change over time.

In the over-15 years since we began the scenes project – and still longer since Terry Clark began studying the urban development potential of amenities like juice bars and opera companies (Lloyd and Clark 2001) – longer time-series of key datasets have become available. We review some of these below while examining various models of scene change. While these data are still not at the level of traditional economic time-series, from a larger historical perspective, we are still in the early days of evolving measures and data sources geared toward investigating the contemporary post-industrial society. This paper accordingly is a first step toward developing theories of how scenes change, and joining these ideas with some data and evidence.

In the spirit of much scenes research, we avoid reducing change to a single dimension but seek multiple overlapping processes. The existing theory of scenes provides guidance to some degree: we seek models for why places (e.g. the physical environment, amenity composition, design) would change; why the people there would change; why what they do would change; and why their values would change. Fortunately, there is a rich tradition of change theories in the history of social thought and research to draw on. As with much of our work, what we offer is not necessarily new (what is?). Rather, joining various theories of change with each other and with appropriate data adds a distinct holistic perspective to change that corresponds to our holistic and contextual approach to examining scenes statically.

In what follows, we elaborate four models of how scenes change. These are the "4 D's" of scene change: development, differentiation, defense, and diffusion. After collating some recent patterns of change in US scenes and amenities, we articulate core concepts for each change model. We then offer initial results demonstrating how to study each model empirically, discussing data and methods where appropriate throughout. The major aim is to indicate potential lines of research to study how and why scenes change, and to suggest the types of data sources and analyses that can advance the theory of scene change.

**The Scenes They are a-Changin': Trends in Amenities and Scenes**

In *Scenescapes*, and much of our published work, we have primarily relied on a single time point. But the trends are often dramatic. The growth of cafes in Seoul is one striking example. As the U.S. Census Bureau has digitized its "zip code business patterns" database (referred to as BIZZIP in *Scenescapes*) and made annual reports more easily accessible, it has become possible to chart overall trajectories of hundreds of types of establishments, many of which have clear cultural significance in defining local scenes.

Figure 1 shows trends in selected amenities from 1998-2016, in New York, Chicago, and Los Angeles, as well as the national average.

**Figure 1. Trends in selected U.S. amenities, 1998-2016**

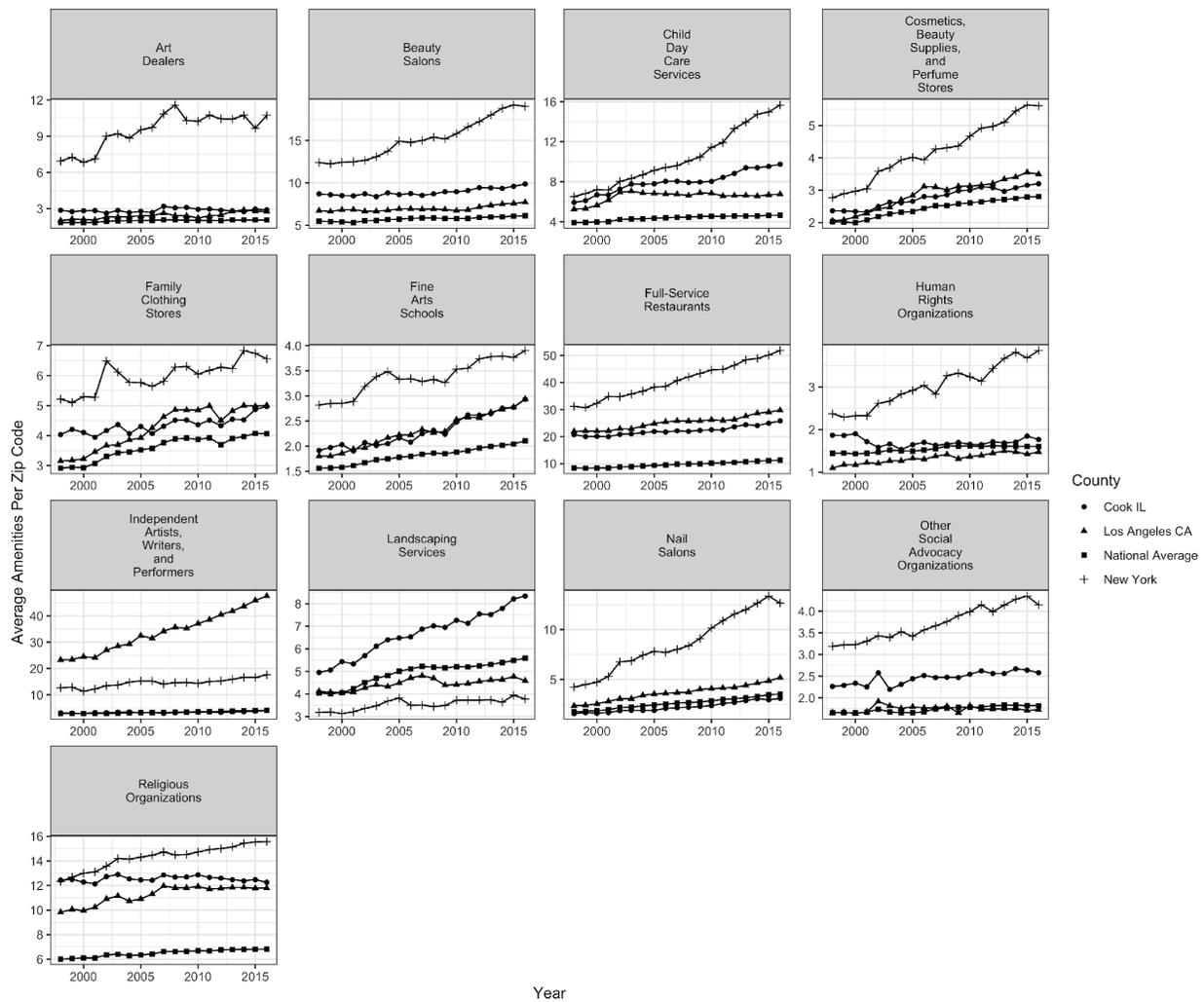

Figure 1 shows dramatic growth in art dealers in New York, in contrast to slow but steady growth elsewhere, along with substantial nationwide growth in fine arts schools. Cosmetic and beauty supply stories, as well as nail salons, have grown sharply across the United States. The average number of restaurants per zip code increased by over twofold. Services devoted to the aesthetics of home and garden grew sharply, with landscaping increasing from 4 to 5.5 nationally per zip code, and to over 8 per Chicago zip code. Children and family-related amenities also grew more prominent, in the form of daycare and family clothing stores. And despite the ongoing trend toward secularization, religious organizations grew steadily, including in cosmopolitan New York City. Los Angeles enjoyed sharp increases in independent artists and human rights organizations, while social advocacy organizations grew especially in New York and Chicago. Overall, these trends in recent decades clearly indicate that cultural, aesthetic, design, educational, sports, and recreation amenities and activities have expanded and entrenched themselves in American communities. The landscape increasingly becomes a scenescape.

Figure 2 adds trends in selected dimensions of scenes. These are z-scores, meaning they show scores in these cities relative to the national average.

**Figure 2. Trends in selected scene dimensions, 1998-2016.**

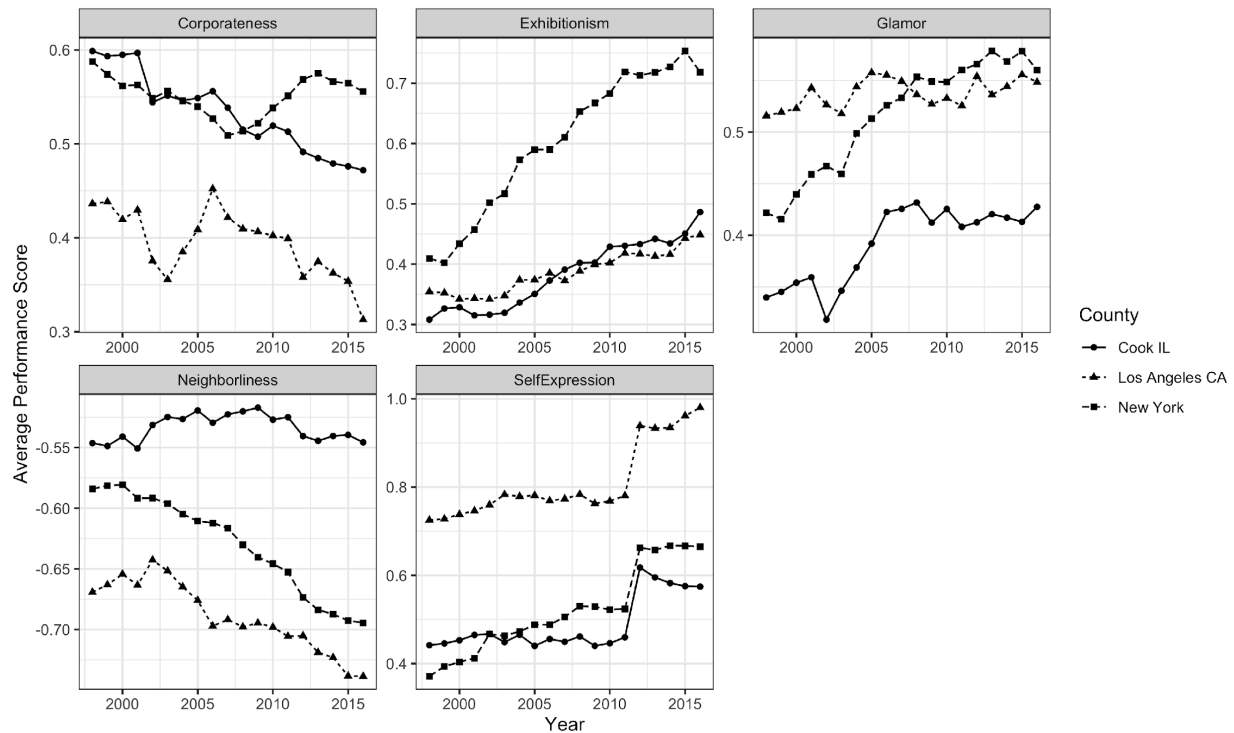

The central messages of Figure 2 are twofold. First, these trends in scenes dimensions help to summarize the overall meaning of changes in not just the handful of amenities shown in Figure 2, but some 150 amenities taken together. While specifics of individual amenities are important to consider, the combined meaning becomes evident in the scenes measures.

Second, and more specifically, Figure 2 shows both continuity and change in the three cities, separately and as a group. Los Angeles retains throughout its distinct status as place that fosters personal self-expression (Silver 2012), though all three grew especially after 2010; similarly, Chicago stands out among major American cities in sustaining a relatively strong sense of neighborliness, while this even declined somewhat in New York and Los Angeles. At the same time, exhibitionistic and glamorous scenes grew in all three cities, but more so in New York, which recently surpassed Los Angeles in overall glamorousness. Corporateness, while high relative to other parts of the country, dropped overall in Los Angeles and Chicago, even as it rebounded in New York in the period after 2007.

Maps reveal the geographic distribution of these trends. Figure 3 illustrates with self-expression, showing for each zip code the difference between its 2016 and 1998 self-expression performance score. In New York, growth in self-expression was highest in the Williamsburg area of Brooklyn, which saw large influxes of artists in this period, though self-expression also increased in other parts of Brooklyn and upper Manhattan. Central Los Angeles saw the highest growth in that city, focused in burgeoning artist areas such as Echo Park and Silver Lake. Growth of self-expression in Chicago was lower overall, but occurred more in the West Loop, Near North side, and near the South Loop, spanning in many ways Chicago's ethno-racial communities. The largest growth occurred in smaller areas with few amenities that saw a few new arts and culture organizations open.

**Figure 3. Mapping trends in self-expression for the three largest U.S. cities, 1998-2016.**

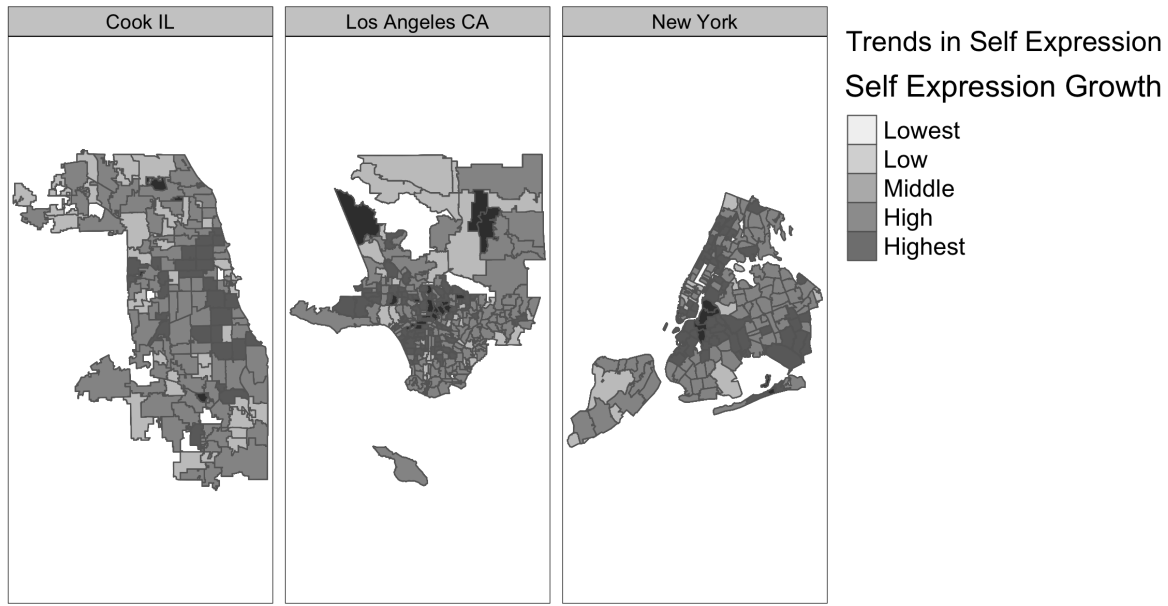

*Note*: this figure shows trends in Self-Expression for all zip codes in New York, Chicago, and Los Angeles. Change is calculated as the difference between self-expression performance scores in 1998 and 2016 (see Silver and Clark 2016). Zip codes are broken into five groups from low to high change, classified by Jenks natural breaks.

### From Description to Explanation: Development, Differentiation, Defense, Diffusion

These and similar trends have analytical value because they enable comparison. Without the ability to compare trends across cities and to the national average, we would have little idea what we were seeing. The meaning of up, down, high, low, consistent, erratic: these are all comparative. We cannot see what is right in front of us unless we have the means to compare it to elsewhere, before, and after. This is a classic idea in social theory from Adam Smith and Charles Cooley to Emile Durkheim. Hence scenes research often insists on the importance of collecting data with a view to broad-based comparisons. The goal is not to be big for the sake of being big. It is because comparison is the beginning of wisdom and self-understanding.

Charting comparative trends is a first step, but it is only a first step. We need to develop theories of change. To do so, we do not need to reinvent the wheel, but can profit from adapting general sociological models of change to the case of scenes. Consider briefly a few such models, and their application to some key new and old scenes data sources.

*Development*. A first core model of change is developmental. This is the simple conception of modernization as a movement from "traditional" to "modern." Ferdinand Tönnies is the classic representative of the idea that as societies develop, they become less traditionalistic and communitarian and more urbane, rationalistic, glamorous, and self-expressive. The emblem of this change is the small tight-knit community and the large city, the former marked by deep tradition and a strong sense of neighborliness, the latter where large corporations, scientific research centers, fashion, and the arts predominate.

More recent authors like Abraham Maslow and Ronald Inglehart continue in this tradition. Maslow emphasized a more or less linear pathway of personal development that follows broadly socio-cultural changes, moving from basic needs like nutrition and personal safety up to social needs such as companionship and recognition and finally cultural needs for self-realization in practices such as art or religion. Inglehart brought similar insights to the longitudinal cross-national research of the World Values Survey, finding that as nations increase in income, education, safety, and security, they tend to see a shift away from "traditional" values and towards "post-materialistic" values of individualism and personal self-expression.

The development model makes a straightforward prediction for scenes. As an area becomes more developed – whether in terms of wealth or education – we should see a shift in scenes dimensions, from those that emphasize communitarian values such as tradition or neighborliness to those that emphasize urbane values such as glamor, rationalism, or self-expression. Call this the **Tönnies-Maslow-Inglehart Theorem**: *where personal and social capacities for thought and action expand, there will be a corresponding shift toward post-materialist values and scenes*.

To test this proposition empirically, we use the same time-series BIZZIP data represented in Figures 1-3. Joining these data with census of population information allows us to conduct statistical tests that were not possible with the single-year dataset used in *Scenescapes* and related work. In particular, we can perform a fixed effects analysis.[1] The power of such models in the present context is that they allow us to examine changes within a given zip code, not only differences between zip codes. With only one time point, only the latter is possible; and while much can be learned from comparing units (such as zip codes), it can also be misleading. If we find for example a positive correlation between higher education and more self-expressive scenes, we will not be sure if that correlation is due to education increasing self-expression or some other feature of highly-educated areas we are not able to observe (it could be that "sophisticated" urbanites both tend to receive more education and value self-expression). While we can gather data to account for as many of these potentially "omitted variables" as we can think of (such as density or proximity to central business districts), we will always leave something out (such as "sophistication").

Fixed effects analysis helps to mitigate this problem. Rather than comparing, say, a downtown area of Chicago or New York to a suburban or small-town neighborhood, instead we compare each area to itself at multiple time points. This means we examine whether an increase of income or education *within* each zip code yields a corresponding increase in self-expression, glamor, or rationalism. In effect, each area becomes a control case for itself. Whatever features of that area remain stable over time can be ruled out as sources of omitted variable bias, even if we cannot directly measure them (such as "urban sophistication"). While this method, like all methods, has its limitations (it does not rule out omitted variables capturing changes within an area), it does constitute a more rigorous way to examine the impacts of changes such as growing income or education.

---

[1] Numerous introductions to fixed-effects models are available. A helpful plain-language summary is here: https://www.kellogg.northwestern.edu/faculty/dranove/htm/dranove/coursepages/Mgmt%20469/Fixed%20Effects%20Models.pdf

Table 1 shows results of six fixed-effects regression models. Each examines the impact of education and income on a scenes dimension: three "urbane" dimensions (self-expression, glamor, and rationalism) and three "communitarian" dimensions (tradition, neighborliness, and egalitarianism).

|  | **Self-expression** | **Glamour** | **Rationalism** | **Tradition** | **Neighborliness** | **Egalitarianism** |
|---|---|---|---|---|---|---|
| *Predictors* | *Estimates* | *Estimates* | *Estimates* | *Estimates* | *Estimates* | *Estimates* |
| Percent with BA or above | 0.01 *** | 0.01 ** | 0.00 ** | -0.00 *** | -0.01 ** | -0.01 *** |
| Median Household Income | 0.01 *** | 0.01 ** | 0.00 ** | -0.01 *** | -0.00 ** | -0.01 *** |
| Observations (zip codes) | 47255 | 47255 | 47255 | 47255 | 47255 | 47255 |
|  |  |  |  |  | *  p<0.05   ** p<0.01   *** p<0.001* | |

**Note**: this table shows results of six fixed-effects regression models at the zip code level. Each examines the impact of education and income on a scenes dimension: three "urbane" dimensions (self-expression, glamor, and rationalism) and three "communitarian" dimensions (tradition, neighborliness, and egalitarianism). All variables have been standardized, and standard errors are clustered by zip codes. Education and income are measured at 2000, 2010, and 2017 and matched to scenes performance scores for each of those years. Scenes data come from Zip Codes Business Patterns, and have been transformed into "performance scores" according to the methods described in Silver and Clark 2016. Census of population data come from The National Historical Geographic Information System.

Results in Table 1 are clear. They provide direct support for the Tönnies-Maslow-Inglehart Theorem. On average, if a zip code experiences growth in education and income, it will also tend to see growth in the urbane dimensions of scenes: self-expression, glamour, and rationalism. At the same time, it will tend to see decline in the communitarian dimensions: tradition, neighborliness, and egalitarianism. These to be sure are simple models with only a few variables included to make interpretation more straightforward. Adding more variables could strengthen the analysis.[2] Even so, this analysis provides the first and most rigorous test to date of factors impacting changes in scenes dimensions over time.

*Differentiation*. Development models of the sort advanced by Tönnies, Maslow, and Inglehart are admirable for their clarity, but have also been subject to sharp criticism for their simplicity. Talcott Parsons criticized Tönnies for assuming multiple analytically distinguishable dimensions would necessarily move in the same direction, even though it is perfectly conceivable that for instance "rationalism" could increase along with an orientation toward serving public goods (as in the case of modern medicine). Maslow and Inglehart have been similarly criticized, for tending to assume a linear trend that always moves away from "traditional" and "basic" needs and values and towards the post-material (Silver 2006). However, a movement toward greater opportunities for self-expression need not reduce traditional social and cultural forms to zero (such as traditional religions) but rather may result in some new equilibrium, as well as novel combinations of tradition and innovation (Silver, Clark, and Graziul 2011). Rather than treat these and other critiques as grounds to abandon development models, they represent opportunities for formulating additional and more refined propositions, which can be tested with scenes measures and data. The above analysis therefore represents a first step and illustration for how to move forward.

---

[2] For example, we explored race as a control with available for two out of the three years corresponding to the education and income data, and found that including it in these models did not alter the results. We plan to incorporate the full "core" model used in *Scenescapes* in the future.

Development models of whatever variety, however, tend to examine relationships between some social, economic, or psychological variables (e.g. income and education) and shifts in substantive cultural *content* or values (e.g. self-expression, tradition, rationalism). Differentiation models of change, by contrast, tend to feature *structural* transformations in patterns or distributions of activities. This is the classic mode of change featured by Emile Durkheim in *The Division of Labor in Society*. Imagine the following highly simplified scenario. At time 1, two towns separated by a river could each have a bakery offering general baking services, and neither bakery would compete with the other. Then a bridge is built between them, enabling movement between the towns. This increases what Durkheim referred to as the "dynamic density" of the area. The bakeries now compete. Either one succeeds and the other goes out of business, or they specialize. We now have a cupcake shop and a cookie shop.

This is of course a highly stylized example, but the more general idea is clear. Call it the **Durkheim thesis**: *communication and interaction engender competition which can be resolved through specialization*. Where scenes evolve in this way, we would expect to find a pattern of increasing specialization, which should be greater in areas with more dynamic density.

We use data gathered from Yelp.com to examine the Durkheim thesis. Yelp.com is a review website in which users can evaluate hundreds of types of establishments, employing a comprehensive directory of organizations. Yelp makes available a dataset for researchers, which includes six cities in the United States and Canada (Toronto, Calgary, Montreal, Pittsburgh, Charlotte, and Phoenix). Every review is time stamped and geo-coded to the corresponding establishment (Olson et al. 2021), which allows us to estimate the number of types of establishments in a given area over time (in the present context, census tracts from 2011 to 2017).

Key for the present analysis is the fact that establishments in Yelp are classified according to a hierarchy that runs from more general to more specialized categories – for example, from the more general category "restaurant" to more specific types of cuisine. Thus we may measure Durkheimian specialization in an area as the relative prevalence of more rather than less fine-grained categories. A less specialized area may have two "restaurants," while a more specialized area could have a French restaurant and a Chinese restaurant.

We translate this intuition into a quantitative measure as follows. The deeper the category is in the yelp hierarchy, the more specialized. The most general level of the hierarchy receives a specialization score of 1; a category one level deeper into the hierarchy receives a score of 2, and so on. In the Yelp hierarchy of categories, there are four levels. Using this weighting scheme, we identify all categories available within each tract, and we sum all their specialization weights. A given establishment may have more than one category. We then normalize this sum by the total number of categories in the tract for each year. This makes the resulting scores comparable across areas with different total numbers of categories. For instance, consider hypothetical tracts A and B. Tract A has 5 categories, each one with a specialization score of 1 (cat1=1, cat2=1, cat3=1, cat4=1, and cat5=1) and Tract B with two categories, one with specialization score of 3, the other with a score of 2 (cat6=3, cat7=2). In this case, the sum with no normalization would be 5 for each tract; however, with the normalizing process, Tract A =1 and Tract B = 2.5. The resulting score reflects the higher average specialization of Tract B.

Figure 4 shows annual trends in specialization for all cities in the dataset, while Figure 5 shows the same trends broken out for each city.

**Figure 4. Overall trends in Yelp.com amenity specialization across six U.S. and Canadian cities**

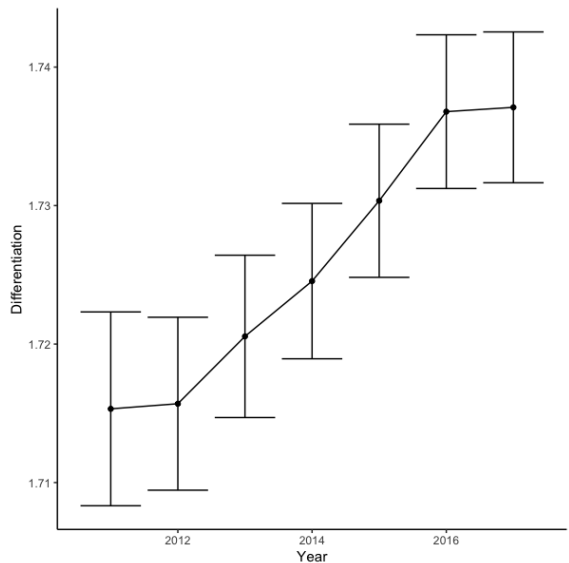

The overall trend is clear. Specialization is increasing over time, though it is slightly decelerating in some places. Overall, moreover, U.S. cities tend to be more specialized than Canadian, and to be experiencing more rapid specialization.

The Durkheim thesis predicts that dynamic density should drive these trends. To examine this proposition, we again turn to fixed effects models. Our primary measure of dynamic density is business density: the total number of establishments per square meter in a census tract. While we also examine population density, business density is more in line with Durkheim's original idea, as in the example of the cupcake shop and the bakery. Business density can also be calculated annually, whereas at the time of writing only a limited number of population variables could be incorporated from more than one time point. In the U.S., these include race and population density, and in Canada, race, population density, median household income, and education.

**Figure 5. Overall trends in Yelp.com amenity specialization within six U.S. and Canadian cities**

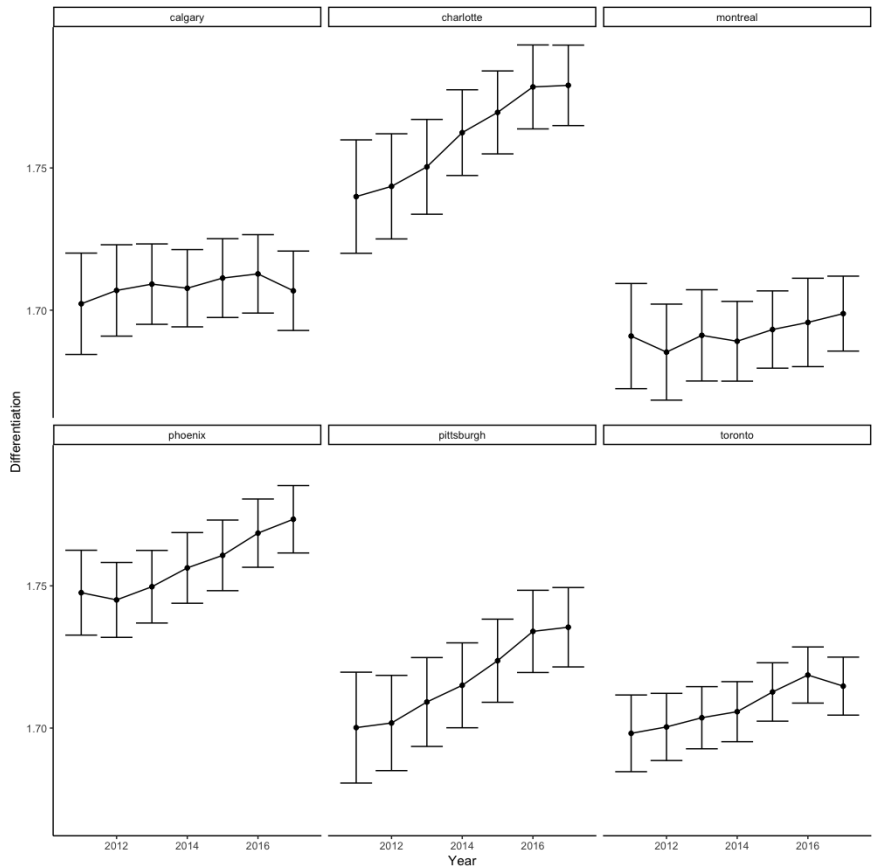

*Note*: figures 4 and 5 shows trends in the degree of specialization in Yelp.com categories in six cities. Specialization is measured in terms categories' position within the Yelp.com hierarchy of classifications, which moves from more to less general categories (e.g. from "restaurant" to "poke restaurant." Areas where the average category is more refined are considered to be more specialized. Points show the annual average, and whiskers show standard errors.

Table 2 shows results for three models: US only, Canada only, and then both countries together. The latter includes only business density, since this allows us to examine the impact across all years in all locations.

**Table 2. Dynamic density causes increasing differentiation.**

|  | Canada | USA | Combined |
|---|---|---|---|
| *Predictors* | *Estimates* | *Estimates* | *Estimates* |
| Percent with BA or above | -0.01 |  |  |
| Population Density | 0.01 | 0.00 |  |
| Median Household Income | 0.01 | | |
| Percent White | -0.00 | 0.01 | |
| Business Density | 0.02 ** | 0.01 * | 0.01 ** |
| Observations | 1265 | 1343 | 9241 |
| | * $p<0.05$   ** $p<0.01$   *** $p<0.001$ | | |

*Note*: Table 2 shows results of three fixed effects regression analyses. In each model, Yelp specialization (measured as described in the text) is the dependent variable. Business density is measured annually; in Canada education, population density, median household income, and race is measured at two time points; in the USA, only population density and race variables could be included at two time points at the time of research.

The results in Table 2 strongly support the classic Durkheim thesis. In both countries, the *only* variable with a statistically significant effect on differentiation is business density. Neither race, nor education, nor income, nor population density show a reliable relationship with change in specialization within Canada or the USA. Moreover, in the combined model covering all years, tracts that increase in business density also tend to experience corresponding increases in differentiation. The Durkheim thesis appears to provide a strong explanation for the growth in differentiation observed in Figures 4 and 5.

*Diffusion*. The differentiation model highlights structural changes in the form of local amenities, from less to more specialized. It does not say anything direct about the substantive value-contents that might correspond to these changes. Classical theorists such as Georg Simmel suggested that greater differentiation should foster more individualism and heightened personal self-expression, while at the same time potentially creating the basis for traditionalistic counter-movements. Having formulated the development and differentiation models, these and similar propositions become viable and tractable topics for future empirical research by joining them in a more interactive model.

Still, development and diffusion do not exhaust the classic models of social evolution. A third model of change is diffusion. Classic theorists of diffusion include Gabriel Tarde, along with later and more recent efforts by the likes of Edward Shils, Paul DiMaggio and Woody Powell, among others. The basic idea in a diffusion model is that ideas and values travel from one place to another, bringing with them social changes. They are often adapted in some way based on how they interact with the local environment. More recent research traditions in the diffusion of innovations (Rogers, Singhal, and Quinlan 2014) and urban evolution carry these ideas forward to the present (Silver, Fox, and Adler, n.d.).

Diffusion can occur through several mechanisms. One is migration: people travel and bring with them their tastes. They create or inspire the creation of scenes on the model of those tastes, again modified in new settings. Daniel Elazar moved in this direction, showing how the American cultural matrix grew out of the migration patterns of European groups with different religious and cultural ideals. Another mechanism is copying, or mimesis: people in one place read about or see on the internet or watch movies about styles from elsewhere, and try to create their own versions. Tarde emphasized this process. Copying can happen from the bottom up or the top down: bottom up when ordinary persons start trying out new styles they learned from elsewhere; top down when government officials or managers impose those styles.

Depending on the type of diffusion, we might expect to see different adoption curves as scenes spread from one place to another. If diffusion is driven by migration, we should see a close connection between the movement of peoples and then changes in scenes occurring in their wake. If it is bottom-up mimesis, we should expect to see an S-shaped curve. Call this the **Tarde Thesis**: *diffusion driven by copying ideas and practices tends to begin slowly, rapidly accelerate at a critical threshold, and then level off.* If it is top-down imposition, we should expect to more of a C-curve that rises rapidly and then levels off. In this process, a central authority dictates the adoption of a practice by certain areas at the same time, generating rapid adoption early-on in the process. Subsequent diffusion is similarly centralized and 'lumpy', directed to designated areas at an appointed time. While S-shaped processes feature organic/emergent diffusion, C-shape ones are associated with design and control.

Figure 6 illustrates idealized forms these core diffusion dynamics can take.

**Figure 6. Idealized diffusion curves.**

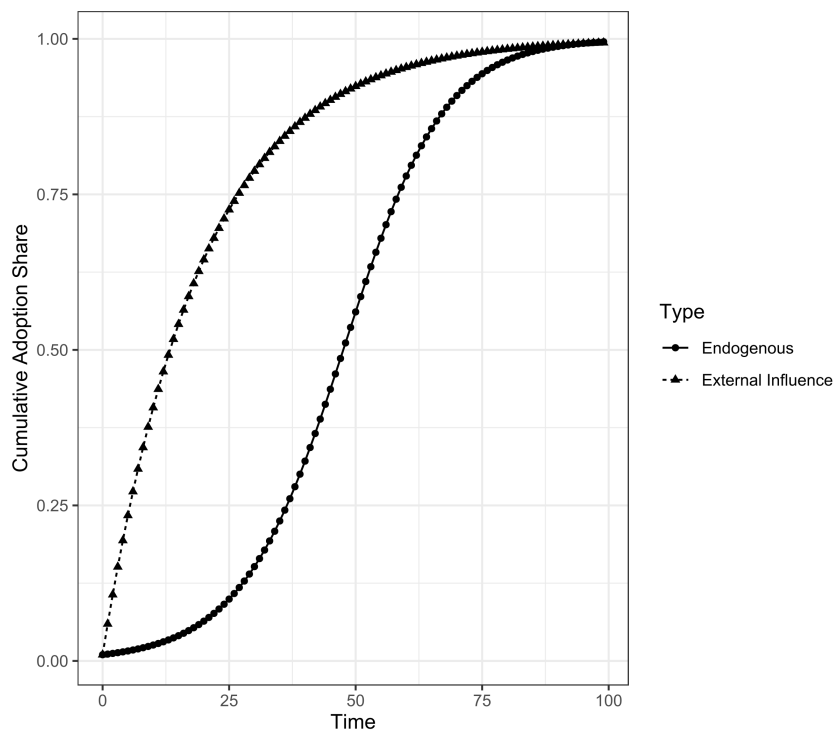

*Note*: Idealized innovation diffusion curves showing a 'C shaped' curve which indicates an external source of diffusion, such as a planning authority, and an S shaped' curve, where the rate of adoption is determined endogenously by a network of adopters responding to informational cues. The latter is associated with fads, chain migration and other forms of social learning.

Ongoing research in the diffusion of innovations and the evolution of urban forms adds ideas about the underlying variables that might affect the specific trajectory of an innovation. For example, (Raimbault and Pumain 2021), borrowing from Hagerstand (1968) argue that innovations tend to start in urban centers before filtering through the urban system to smaller towns. Other research points to imprinting, whereby practices become stamped by the values and reputations of their origins, which in turn limits their spread to similar locations until or unless some adaption renders it acceptable to new groups. For example, yoga initially entered the USA via highly educated college towns such as Cambridge, MA or Berkeley, CA, and its subsequent spread has been restricted to similar areas (which themselves have expanded). For yoga to make an "evolutionary leap" in America, it would have to adapt to appeal to areas outside of its current niche (Yi and Silver 2015).

Figure 7 shows diffusion patterns for four kinds of organizational forms, hinting at how diffusion curves can be used to understand underlying processes: Blaze Pizzeria (a North American Pizza franchise), Tesla Supercharging Stations, Latter Day Saints Churches (Mormon), and countries in the European Union. For each form, the figure shows the total percentage of diffusion by year of real time. In the cases of Blaze and Tesla, forms that tend to be popularized through social learning (individuals or organizations observing others and reproducing their behaviors or tastes), we see rapid expansion in a matter of a few years. The European Union diffuses to more countries in a more methodical way, via waves of expansion that are already known to be coordinated at the Union and country levels. In between, we see LDS churches, which saw their own exponential growth in the first thirty years, as the religion was popularized, followed by waves of expansion once the church had acquired centralized planning functions. Such a subtle, 'hybrid' pattern is common with real world data and can be a powerful clue into the evolutionary pressures that are acting on urban forms and infrastructures at given points in time.

**Figure 7. Diffusion curves for four types of urban forms.**

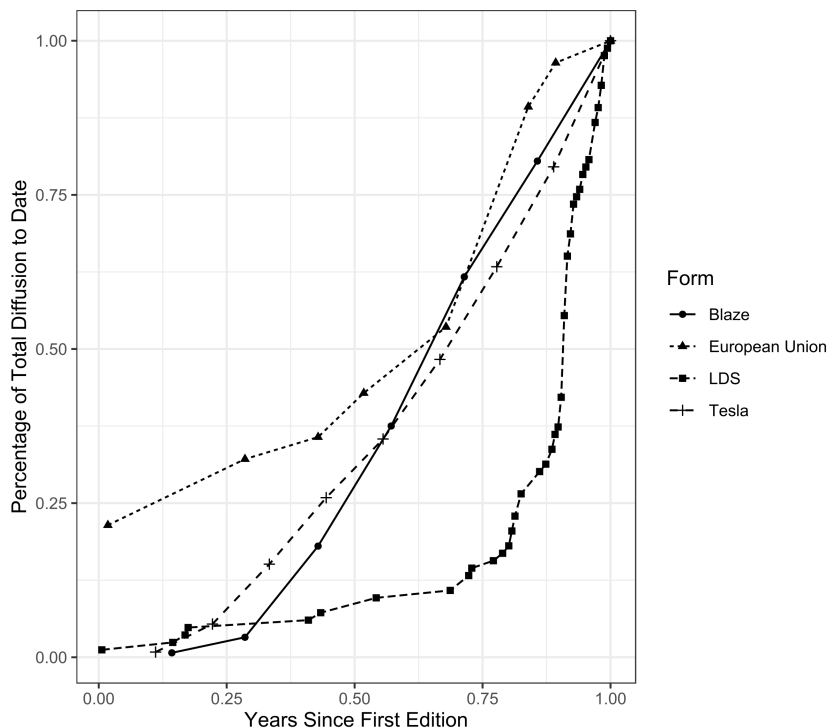

*Note*: Figure 7 shows the pace of diffusion for four different forms: Blaze Pizza Franchises (taken from the store website), European Union Countries, LDS Churches (Taken from the Church Website) and Tesla Supercharging Stations (taken from supercharge.info). All data is current as of 2019. The X axis shows the number of years since a form first appeared and the Y Axis shows the proportion of 2019 editions that had diffused by that time.

These specific amenities (and the others discussed below) have time-series data available showing when and where each new location appeared from their origination, making them good candidates to examine diffusion. While they of course are not direct measures of scenes, their adoption clearly has implications for the scenes in which they do or do not appear. What Figure 7 specifically indicates is the empirical difference between relatively top-down versus bottom-up diffusion processes.

Figures 8 and 9 illustrate how diffusion trajectories can be inflected by the characteristics of the origin and destination of a given practice.

**Figure 8. Size distributions of Apple Stores locations according to their diffusion trajectories.**

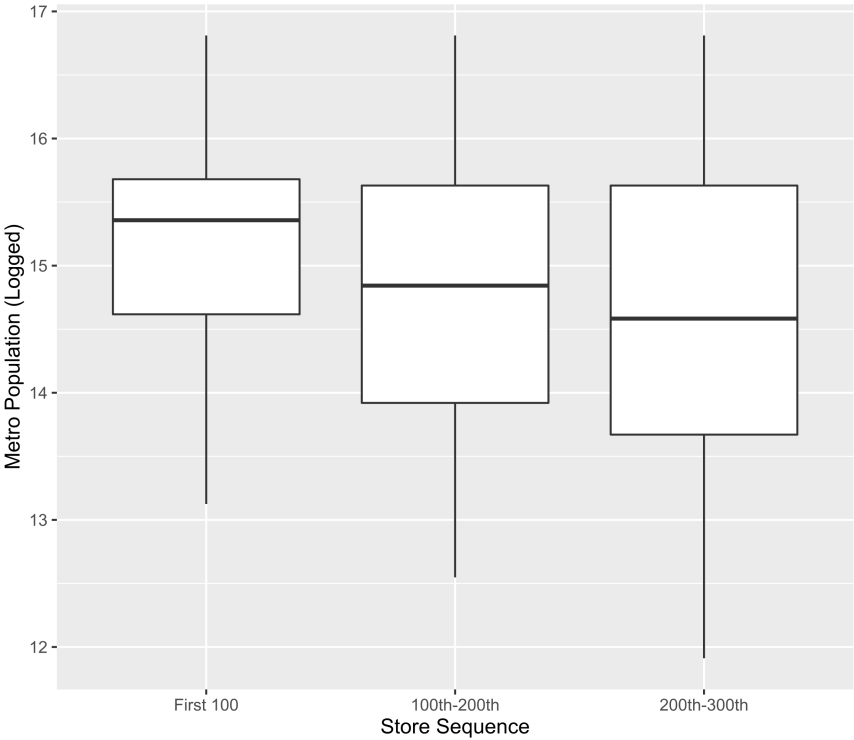

*Note*: Figure 8 summarizes the size distributions of the metro regions containing the first 100, Second 100 and Third 100 US-based Apple Stores. Data is from the Apple Store website as of 2018. Metro size data is from the 2015 American Community Survey.

**Figure 9. Education level of Walmart levels according to their diffusion trajectories.**

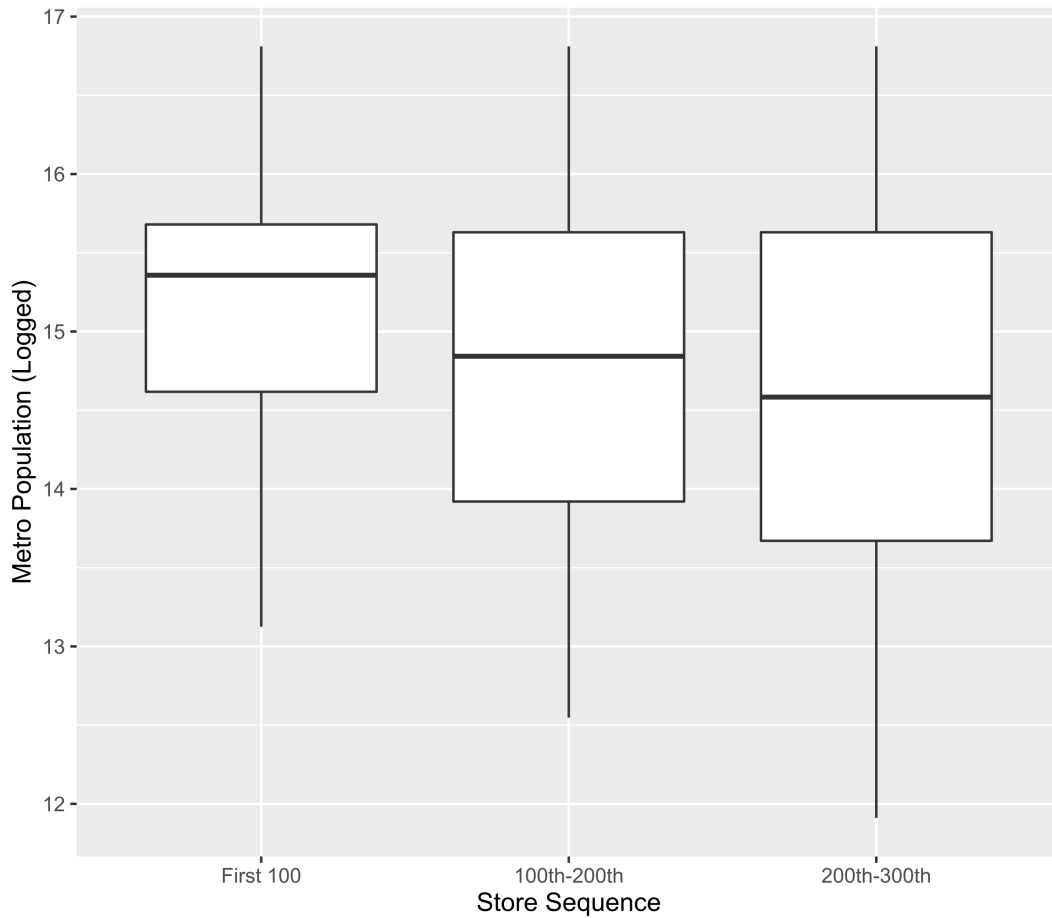

*Note*: Figure 9 summarizes the average degree attainment of adults in metro regions containing the first 100, next 900 and next 2165 US-based Walmart Stores. Data is from AgData as of 2018. Education data is from the 2015 American Community Survey.

Figure 8 follows the trajectory of Apple stores from their early to later locations. The first 100 Apple stores tended to be located in larger metro areas, the second 100 stores in somewhat smaller metro areas, and then the most stores in year smaller metro areas. This is a fairly clear example of the sort of filtering down the urban size hierarchy hypothesized by Pumain et al. The right panel adds more specific zip code information, however, in this case to the diffusion of Walmart stores. Walmart in the USA has strong cultural associations, often with lower educated rural areas and their characteristic tastes. Some election forecasters even use Walmart proximity to predict where Republic vote shares will tend to be higher. Figure 9, however, complicates this picture. It shows that for its first 1000 stores, Walmart did indeed tend to locate in zip codes with less education, but especially the more recent stores have been in increasingly educated areas.

This is both an illustration of the potential for some practices to undergo evolutionary adaptations that open them to new ecological niches and of the tendency of Walmart and similar superstores to advance what (Yi and Silver 2015)refer to as the distinctive pragmatist strand of the American cultural matrix. Such pragmatist businesses stock not only symbols of rural American culture (traditional religious items, hunting equipment) but also more expressive and urbane items characteristic of more educated areas (yoga supplies, organic groceries). In this way, the diffusion pattern of Walmart shows its distinctive place in bridging across American cultural divides. Additional research can benefit by examining what types of activities tend to remain relatively restricted to a specific scene in contrast to those that over time are able to span many types. Preliminary research indicates for example that amenities like Tesla charging stations and In N Out Burgers are spreading from less to more traditionalistic scenes, while Walmart stores are doing the opposite. This is an area ripe for further research.

Figure 10 provides a final illustration of diffusion processes, this time focusing on examining travel and taste patterns using social media data (in this case, check-ins on Foursquare).

**Figure 10. Behavioral distance between tourists and locals on Foursquare.**

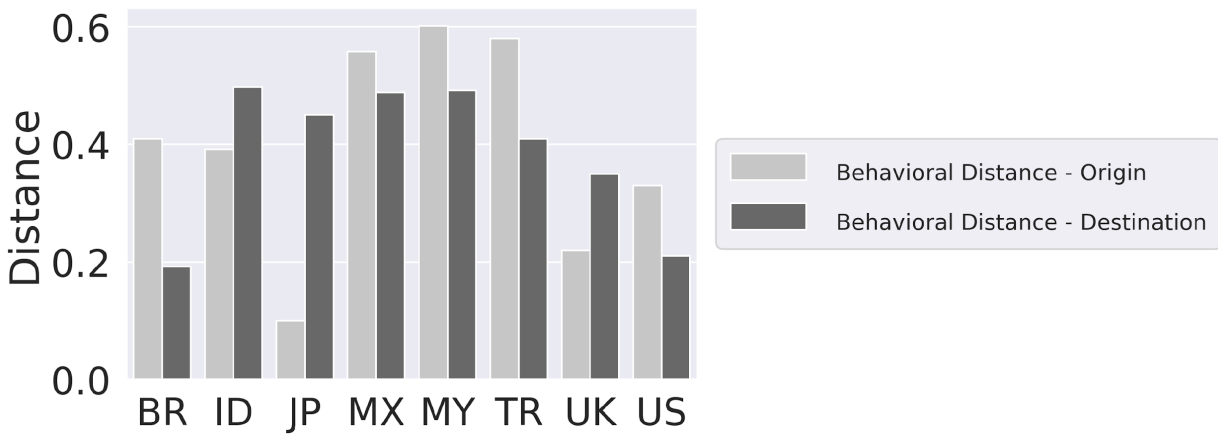

Figure 10 is based on Foursquare check in activity patterns. On Foursquare, users check in to various amenities and points of interest they visit. As they do this more and more, each user acquires a characteristic taste profile: the range of restaurants and other venues they prefer. Not only is this a strong indicator of those users' taste for distinctive scenes, it also allows the researcher to examine how those taste profiles change as users travel to new locations, relative to typical users in their home and destination areas.

Silva and colleagues did just this, with a large dataset of Foursquare users from eight countries (Senefonte et al. 2020). Brazilian tourists in particular stand out in that they tend to absorb more intensively the local habits of countries visited, in contrast to tourists from other countries. Japanese tourists by contrast tend to maintain their original taste profile when leaving the country: their activities remain close to Japanese residents, even when they travel to countries very culturally distant from Japan. While much work remains to generalize from these findings, they highlight key processes in the diffusion of scenes, ranging from the tendency of some groups to strongly persist in their local practices as they carry them to new locations, and for others to absorb diverse tastes and bring them back in new combinations to their homelands. Depending on these mechanisms, scenes at home and abroad would likely evolve in distinctive directions.

*Defense*. Diffusion models investigate the movement of peoples, practices, and ideas from place to place, examining the changes that result from their arrival and adaptation in new locations. While existing research has to some extent incorporated cultural characteristics of origin and destination locations (Silva and colleagues for example use the World Values Survey to compute cultural distance measures at the national level), adding additional local scenes measures as discussed above promises to refine and deepen existing diffusion research. Moreover, as social media data expand, it becomes increasingly possible to measure both the scene of a place and the scene-orientations embodied in people's tastes and behaviors, and to study how the two change (or not) as they interact.

Indeed, existing scenes often do *not* change when newcomers with new ideas and tastes arrive on the scene. Representatives of the existing scene may resist, seeking to maintain the current style of life. Stability is not a static process: the current state of affairs continues because some people work to keep it that way, against pressures to the contrary. In such situations, scenes change occurs when resistance is overcome or fades away.

We may refer to these processes of defense, challenge, resistance, and overcoming as the defense model of change. In this model, existing types of scenes tend to maintain themselves via various practices that serve to resist potential changes. When new types of activities become too strong or distinct to be counteracted, the scene changes. For example, when locals protest against newcomers in order to maintain the local authenticity of their scene, this is an example of activities that function to sustain the existing structure.

This model sometimes overlaps with conflict theories, as articulated by Arthur Stinchcombe, and embodies what can be called a "functionalist" logic. Functional explanations, broadly speaking, regard some consequences of a social arrangement or activity as crucial causes of the arrangement or activity in question (Stinchcombe 1987). For example, an urban researcher who hypothesizes that gated communities' guard stations exist to maintain the social standing of their residents by excluding members of marginalized groups is proposing a functional explanation. The consequence, "maintaining social standing," is hypothesized to be the cause of the social arrangement, "guard stations."

While functional explanation has predominated social research in the past, it has largely fallen out of favour in social science (Jackson 2002).There are several common criticisms (Merton 1968)). These include charges that functionalist explanations assume overarching goals shared by entire social systems, thereby denying the possibility of social conflict -- *functional unity*; that all social entities, if they exist, must therefore serve some (possibly latent) social function, leaving the researcher free to posit functions even for random or contingent arrangements -- *universal functionalism*; and that some functions, usually served by traditional institutions, are so important that they cannot be altered, fostering unwarranted complacency and conservatism -- *functional indispensability*. These criticisms were widespread through the 1960s and 1970s, and were especially potent when directed against Parsonian structural-functionalism and its anthropological cousins. They persist today, even as functional explanation is less frequently advanced, at least explicitly.

Despite these and other criticisms, functional explanations need not assume society's functional unity, universal functionalism, or functional indispensability. For any proposed functional explanation, one must ask: functional for whom? Conflict is a constant possibility in a functional account. Similarly, universal functionalism is not inherent to functional explanations. When properly specified, a functional hypothesis can be falsified: a purported function may fail to exhibit the expected set of functional relationships (Joas and Knöbl 2009).Suppose empirical observation shows that members of marginalized groups frequently become residents of gated communities. In that case, one might conclude that the guard station does not function to exclude such groups.

Nor must functional accounts promote a conservative view of society. From Marx and Engels' claim that, under capitalism, governments exist to serve the interests of the bourgeoisie, to more recent arguments by Pierre Bourdieu that social connections and cultural attainment do so as well, critical theorists have made strong claims about how the purported functions of various social arrangements perpetuate their existence. For these and similar reasons, recent philosophers and theorists of social science have offered ``qualified defences'' of functional explanation (Joas and Knöbl 2009; Jackson 2002).

Despite its bad reputation, social scientists routinely use functional motifs in their accounts of social life. (Jackson 2002) shows their ubiquity in economic theory, and in a recent review of the literature on neighbourhood change, we found numerous examples there as well (Daniel Silver and Silva 2021). However, researchers rarely articulate such functionalist motifs as such. Nor do they formulate them in terms of a model that would allow their implications to be evaluated empirically: "references to functional relations in the social scientific literature are seldom linked with evidence that these relations do exist" (Joas and Knöbl 2009). In fact, we found no examples in the neighbourhood change literature where an author who utilized a functionalist motif articulated the motif in an explanatory model that would render it testable.

Figure 11 however shows how to do so.

**Figure 11. General model of functional explanation (left) adapted to the case of neighborhood change (right)**

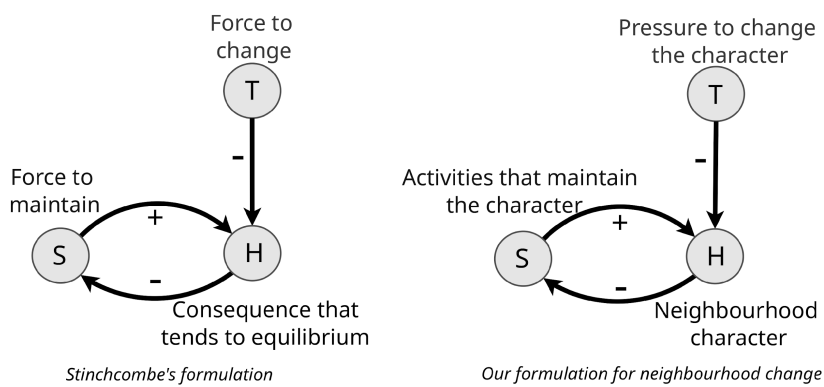

*Stinchcombe's formulation*     *Our formulation for neighbourhood change*

*Note*: this figure shows the core structure of functional explanation in general (left) and applied to local neighborhood and scene change (right).

The left panel of Figure 11 describes the core causal structure of a functional explanation, drawing from Stinchcombe's (1987) seminal work. The diagram is composed of the following analytical elements:

- The consequence that tends to be maintained, which also functions indirectly as a cause of the behaviour or social arrangement to be explained. This is H, the "Homeostatic" variable. Though H may tend to be stable empirically, its stability is maintained against pressures to change it, such as in the case of body temperature.

- The social arrangement or behavior that impacts H, the explanandum. This is S, the "Structure." In a functional model, Structures tend to maintain Homeostasis. For example, sweat glands tend to maintain body temperature.
- Tensions that tend to upset Homeostasis, unless Structures maintain it. This is T, the "tension" variable. If physical activity or air temperature did not alter body temperature, there would likely be no structure to counteract the tensions they create.
- Processes that reinforce or select for the S's (structures) that maintain H (homeostasis). When H is threatened or pressured, these forces increase the activity of S when T (tensions) are higher and decrease when H is maintained. For example, sweat glands generate more sweat (S) when body temperature (H) is not maintained at normal levels due to a certain phenomenon (T). Since this structure helps to maintain H in equilibrium, it will tend to be selected or reinforced.

To illustrate this diagram, Stinchcombe uses the example of magical rituals to reduce anxiety around sea fishing. In that example, the homeostatic variable H is anxiety, the tension variable T is uncertainty in fishing, and the structure variable S is magical activity. As T (uncertainty) increases, so does H (anxiety). However, S (magical activity) compensates and reduces anxiety to tolerable levels. This has the effect of reducing variation in H, maintaining relatively stable anxiety levels over time. Moreover, S (magical activity) becomes selected out and reinforced in proportion to its effectiveness in maintaining tolerable anxiety levels; were other more effective structures (e.g. improved ship design or weather forecasting) to emerge, they would tend to be reinforced instead.

Stinchcombe's diagram may be intuitively mapped onto familiar neighbourhood dynamics, as shown in the right panel of Figure 11. For example, we may treat as Homeostatic (H) variables neighbourhood character, style, or scene (such as distinctive shops, restaurants, venues, or groups), Tension (T) variables as pressures to change that character (from, for example, new groups with divergent tastes), and Structure (S) variables as activities that maintain that character (such as Business Improvement Association sponsored festivals, political advocacy, or increased participation in venues and activities distinctive to that scene).

Based on this model, Silver and Silva (forthcoming) formulate six hypotheses about functional dynamics of scene change. Here we highlight two key propositions:

- *Structural response*. Local increases in Tension (T) should be followed by increases in Structure (S).If neighborhoods function to maintain their existing character, as pressures to change grow, this should be followed by increases in activities that correspond to that character. For example, if new residents threaten the "bohemian" character of a neighbourhood, activities that intensify the bohemian style of the area should increase (e.g. offbeat or transgressive art activities).

- *Significant response*. Increases in Structure activities (S) should be proportional to the strength of Tensions (T) to which they are subjected, i.e., the increase should also be substantial. If neighborhoods function to maintain their existing character, the structural response should not only follow rising Tension (P2); it should be larger when the Tension is larger. For example, the more "square" or "bourgeois" new residents in a "bohemian" area are, the more intense the bohemian response should be.

We test these propositions, again using data from Yelp.com. To do so, similar to the analysis of Foursquare described above, we build taste profiles for all Yelp users (across the same six cities discussed above), based upon the categories of amenities they have reviewed. This profile may be vied as each individual's "scene orientation," though further work could explicitly code such profiles in terms of the 15 dimensions of scenes. We then identify "regulars" for each neighborhood, and "newcomers." This in turn permits us to test the structural response and significant response propositions, by examining the extent to which a) regulars increase their local activities on Yelp after newcomers increase theirs and b) this increased activity is greater when newcomers' activities are also greater, especially when the latter have taste profiles very different from the former.

Figure 12 shows activity levels for both regulars and newcomers in one neighborhood in downtown Toronto.

**Figure 12. Scene defense in a tension-response dynamic**

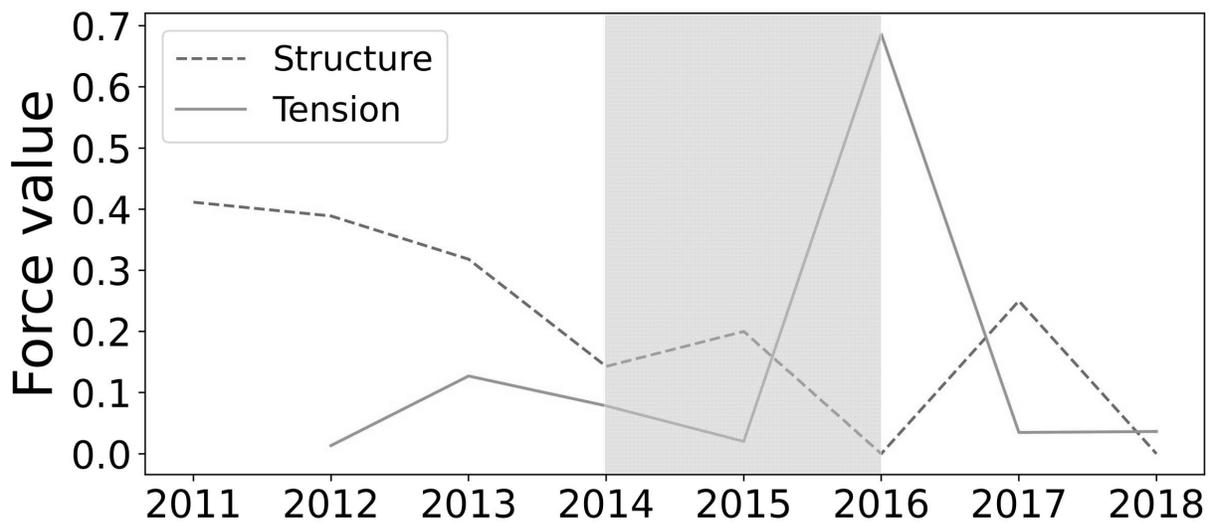

*Note*: this figure shows the tension-response dynamic at work in one census tract in downtown Toronto. The blue line shows how frequently regular visitors to the area use its most popular types of amenities. The orange shows activity from users with taste profiles different from the regular users. The figures shows that when the orange line spikes, the blue line exhibits a corresponding increase afterwards. Additional research shows that this pattern is common across six US and Canadian cities, indicating that the "defense" response to pressures to change the scene is widespread.

The Structure (blue line) represents how frequently regular visitors to the area use its most popular types of venues. In contrast, the Tension (orange line) represents new users with taste profiles distinct from prior users. In this example, in 2016, tension comes from more newcomers who tend to prefer shopping and fashion and fewer newcomers that prefer nightlife and bars. The Structure created by regulars is mainly composed of restaurants, bars, and nightlife in this area. The peaks observed in 2016 for the orange line and 2017 for the blue line stand out from recent trends: an increase in Tension is followed by rising Structure activity, indicating the presence of a functional process.

Further analysis shows that this pattern of tension-response is not unique to this specific neighborhood, but is in fact the norm across all six cities that we studied. Moreover, these increases in responses by regulars yield real changes on the ground in their areas. Where a structural response to challenges by newcomers is stronger, local amenities over time tend to be selected that conform more closely to the regulars' tastes, while those that are more culturally distant from their tastes disappear. By contrast, where the locals are *not* able to mount a strong structural response to challenges, the local scenes tend to change in line with the tastes of the newcomers. Silver and Silva 2021 document these processes in more detail, but the import for scenes research is clear: functional processes of scene maintenance, pressure, challenge, and defense are crucial in determining when and why diffusion of tastes impact an area, or are resisted. Challenge and defense are as much a part of scene change as communication and dialogue.

**Conclusion**

The major goal of this paper has been to articulate and apply multiple highly general models of social change to the more specific processes in changing scenes. After describing some broad trends in U.S. amenities and scenes since 1998, we laid out four broad models of change, the 4 D's of changing the scene: development, differentiation, diffusion, and defense. While these four models are not original to us, elaborating them explicitly and in tandem is an important contribution to sociological theory in itself, as is gathering research and data that allows them to be empirically evaluated.

The four D's help identify changes in key aspects of scenes: values, practices, people, and places. Development features processes that generate changes in values. Classic development hypotheses highlight how "modernization" shifts toward individualism and rationalism and away from traditional and communal values. While we found evidence of such processes, we noted that this is not the only possible form of development. Differentiation highlights changes in places and practices. Classic differentiation hypotheses highlight how "dynamic density" leads to more specialized amenities and activities. Our analysis showed strong support of this old but powerful idea from Durkheim. Diffusion features changes in people and practices. Classic diffusion hypotheses tie social and cultural changes to the movements of people and ideas. Our analysis highlighted how this can occur through top-down and bottom-up processes, issuing in different paces and patterns of change, as well as how diffusion is channeled by the meanings imprinted upon practices in their origin locations, hierarchical filtering among metropolitan areas, and the tendency of people from different places to be more or less resistant or open to changing their tastes as they visit new places. Defense primarily features changes in people, from which other changes follow. Classic defense hypotheses highlight how incumbent locals respond to challenges from newcomers to maintain the existing character of the local scene. If the defense fails, the scene changes.

In sum, the four D's together allow for robust analysis of multiple overlapping mechanisms by which scenes change. Formulating a host of change models moreover advances the holistic spirit of scenes research. With these (and more) models of change in the analytical toolkit, the scenes researcher may apply one or more of them to specific cases where appropriate.
Additional theoretical research may go further by explicitly joining these models.

Together with emerging new data sources, such integrative research promises to open new horizons for understanding how and why scenes change. While the models themselves may extend to international contexts, the specific dynamics should vary by context. For example, Wu and colleagues' (2019) research in Beijing found that self-expression alone does not have a significant effect on the agglomeration of creative occupations and activities. Instead, it is necessary to combine dimensions such as education (number of colleges and universities), willingness to experiment, and a strong focus on rationalism to have a significant effect on the agglomeration of creative workers. They suggest that self-expression is modified in the context of Confucian culture: if one is too assertive or too self-expressive, it will be understood as being less polite and respectful to others.

This is an example of how change processes can be influenced by the broader cultural context. As diffusion, development, differentiation, and defense continue to operate within and between nations, new directions for understanding how and why scenes persist or change open up. For example, Chinese immigrants in Toronto bring aspects of Chinese scenes to Canada, which in turn are adapted and modified in the new context. This and similar change processes are currently being studied by numerous scholars working in dialogue and communication. As international collaboration among scenes researchers widens and deepens, the present study charts a path forward for incorporating analysis of how scenes grow and evolve in the hopes of guiding deeper study of international similarities and differences.